\documentclass[aps,pre,reprint,amssymb]{revtex4-1}

\usepackage{amsmath}
\usepackage{bm}
\usepackage{graphicx}			
\usepackage[colorlinks=true, linkcolor=blue]{hyperref} 

\newcommand{\N}{\mathbb{N}}

\newcommand{\R}{\mathbb{R}}

\renewcommand{\i}{\mathrm{i}}
\newcommand{\hamilton}{\hat{\mathcal{H}}}
\newcommand{\kk}{\frac{k}{\kappa}}

\newcommand{\typpref}{\sqrt{\frac{m}{2\pi \i\hbar t}}}

\newcommand{\tkappa}{\theta^{0}_{1}}
\newcommand{\ktilde}{\tilde{\kappa}}
\newcommand{\mutilde}{\tilde{\mu}}
\renewcommand{\Re}{\operatorname{Re}}

\DeclareMathOperator{\erfc}{erfc}

\newcommand*{\Diff}[2]{\mathop{}\!\mathrm{d}^{#1}{#2}\,}
\newcommand*{\diff}[1]{\mathop{}\!\mathrm{d}{#1}\,}
\newcommand{\invLaplace}[3]{\mathcal{L}_{#1}^{-1}\Big[#2\Big](#3)}
\newcommand{\Laplace}[3]{\mathcal{L}_{#1}\Big[#2\Big](#3)}
\newcommand{\eh}[1]{\mathrm{e}^{#1}}
\newcommand{\vect}[1]{\mathbf{#1}}
\newcommand{\freeprop}[1]{\typpref\eh{-\frac{m}{2\i\hbar t} #1^2}}
\newcommand{\derivative}[2][]{\frac{\partial^{#1}}{\partial {#2}^{#1}}}
\newcommand{\secondderivative}[1]{\frac{\partial^2}{\partial {#1}^2}}

\begin{document}

\title{Semiclassics in a system without classical limit:\\ The few-body spectrum of two interacting bosons in one dimension}

\author{Benjamin Geiger}
\email[Corresponding author: ]{benjamin.geiger@ur.de}
\affiliation{Institut f\"{u}r Theoretische Physik, Universit\"{a}t Regensburg, D-93040 Regensburg, Germany}

\author{Juan-Diego Urbina}
\affiliation{Institut f\"{u}r Theoretische Physik, Universit\"{a}t Regensburg, D-93040 Regensburg, Germany}

\author{Quirin Hummel}
\affiliation{Institut f\"{u}r Theoretische Physik, Universit\"{a}t Regensburg, D-93040 Regensburg, Germany}

\author{Klaus Richter}
\affiliation{Institut f\"{u}r Theoretische Physik, Universit\"{a}t Regensburg, D-93040 Regensburg, Germany}
\date{July 12, 2017}

\begin{abstract}
We present a semiclassical study of the spectrum of a few-body system consisting of two short-range interacting bosonic particles in one dimension, a particular case of a general class of integrable many-body systems where the energy spectrum is given by the solution of algebraic transcendental equations. By an exact mapping between $\delta$-potentials and boundary conditions on the few-body wavefunctions, we are able to extend previous semiclassical results for single-particle systems with mixed boundary conditions to the two-body problem. The semiclassical approach allows us to derive explicit analytical results for the smooth part of the two-body density of states that are in excellent agreement with numerical calculations. It further enables us to include the effect of bound states in the attractive case. Remarkably, for the particular case of two particles in one dimension, the discrete energy levels obtained through a requantization condition of the smooth density of states are essentially in perfect agreement with the exact ones.  
\end{abstract}

\maketitle

\section{Introduction}
The discovery by Bethe of quantum many-body systems admitting analytical expressions for eigenstates and eigenenergies in terms of the implicit solutions of algebraic transcendental equations \cite{bethe1931,lieb1963,yang1969} marked the birth of a whole branch of mathematical physics, namely, the theory of quantum integrable models (for an introduction see \cite{korepin1997}). Since then, these models have served as the playground to study the role of symmetries and conservation laws for the stationary and dynamical properties of many-body systems \cite{gaudin1971,batchelor2005,sykes2007,bolte2017}.

Solvable models had been mainly restricted to the area of mathematical physics because they involve apparently unphysical $\delta$-type interparticle interactions and are commonly restricted to one-dimensional (1D) motion. This situation has drastically changed with the successful preparation of quantum states of interacting cold atoms \cite{ketterle1995,bloch2008}, especially in regimes where the system can be considered essentially 1D, as in elongated optical traps \cite{morizot2006,griesser2013,kim2016}. As it turned out that interactions between neutral cold atoms are described with astonishing precision by $\delta$-type potentials \cite{olshanii1998,chin2010,cazalilla2011,krutitsky2016}, the knowledge accumulated during almost one century of work on solvable models found its way into cold-atom physics during the last decade. Hence, the experimental study of solvable many-body systems has become a very active field that provides hints for other hitherto inaccessible regimes where external potentials destroy integrability.

There are, however, at least two situations where the detailed level-by-level calculation of the many-body spectrum using methods of integrable quantum systems overshoots in the problem of calculating macroscopic properties like the microcanonical partition function. One example is the thermodynamic limit, where quantum fluctuations are suppressed and the spectrum behaves effectively smooth. In this regime the appropriate tool is the so-called thermodynamic Bethe ansatz \cite{yang1969}. The second situation appears when studying interacting bosonic systems in the mesoscopic short-wavelength regime, where a separation of scales between the smooth and oscillatory part of the level density allows for the approximation of the spectrum as a smooth function. This is a well known procedure often referred to as the Thomas-Fermi approximation \cite{brack2003}.

The Thomas-Fermi density of states is of paramount importance as it fixes the energy-dependence of the mean-level spacing, the fundamental energy scale that heralds the appearance of quantum effects and determines the relative importance of external perturbations. In this context, cold atom systems pose an interesting problem: since the Thomas-Fermi approximation explicitly requires a classical limit for the quantum mechanical Hamiltonian, it is not well defined for systems with $\delta$-like interactions where the classical limit only exists away from the collisions.

All in all, in systems with few particles interacting through short-range interactions, the semiclassical limit appears to be difficult to handle already when we ask for a very fundamental quantity as the mean level spacing. Our objective in this paper is to show how, by using techniques imported from the calculation of Thomas-Fermi approximations in single-particle systems with mixed boundary conditions, one can obtain a rigorous definition of the smooth density of states in systems of few particles interacting through $\delta$-potentials, even when the latter do not have a classical limit.

\section{\boldmath One particle in a $\delta$-potential}

In this section we present the general formalism for the calculation of the smooth part of the density of states (DOS) for $d$-dimensional billiards with either $(d-1)$-dimensional $\delta$-barriers [i.e.\@ a $\delta$-function potential along a ($d-1$)-dimensional manifold] inside the volume or with Robin (or mixed) boundary conditions on the surface. The latter case was already discussed by Balian and Bloch \cite{balian1970,balian1971} and by Sieber \cite{sieber1995}, but in all three cases the derivations were done via the energy dependent Greens function whereas we use a formalism based on 1D propagators \cite{baltes1976,hummel2013}. For this we need the propagator for a $\delta$-potential in 1D, which is known exactly from a path integral calculation \cite{bauch1985} (see also \cite{grosche1994} for higher dimensions). Notably, quantum mechanical properties, like the appearance of a bound state for attractive interaction, are well hidden inside the results (see e.g.\@ \cite{goovaerts1973}). Therefore we take an alternative approach which is more instructive for our purposes.
\subsection{\boldmath Propagator for a particle in a $\delta$-potential\label{subsec:propagator}}
	The 1D-propagator for the $\delta$-barrier is derived in a straightforward way by first calculating the exact propagator for a particle on a line with Dirichlet boundary conditions at $x=\pm L$ and a $\delta$-potential $V(x)=(\hbar^2\kappa/m)\,\delta(x)$ with $\kappa\in\R$ and then taking the limit $L\to\infty$. The solutions of the stationary Schr{\"o}dinger equation
		\begin{equation}
			\left(-\frac{\hbar^2}{2m}\secondderivative{x}+\frac{\hbar^2\kappa}{m}\delta(x)\right)\psi(x)=E\psi(x)
			\label{eqn:schroedinger_eq}
		\end{equation}
	for the confined system are well known (see e.g.\@ \cite{pedram2010}) and can be separated into symmetric and antisymmetric solutions due to the symmetry of the problem. The latter are not affected by the $\delta$-potential and coincide with the antisymmetric solutions of a particle in a box:
		\begin{equation}
			\psi_n^{(\mathrm{a})}(x)=\frac{1}{\sqrt{L}}\sin(k_n^{(\mathrm{a})} x),\qquad k_n^{(\mathrm{a})}=\frac{n\pi}{L},\quad n\in\N.
			\label{eqn:antisymmetric_solutions}
		\end{equation}
	The symmetric solutions are given by
		\begin{equation}
			\begin{split}
			\psi_n^{(\mathrm{s})}(x)&=\frac{A_n}{\sqrt{L}}\sin\bm{(}k_n^{(\mathrm{s})}(|x|-L)\bm{)},
			\\
			k_n^{(\mathrm{s})} L&=n\pi-\arctan\!\bigg(\frac{k_n^{(\mathrm{s})} }{\kappa}\bigg),
			\end{split}
			\label{eqn:symmetric_solutions}
		\end{equation}
	where $A_n$ is a normalization constant that depends on $k_n^{(\mathrm{s})}$. The transcendental equation in Eq.\@ (\ref{eqn:symmetric_solutions}) has exactly one solution for every $n\in\N$ and another nontrivial solution for $n=0$ if $\kappa$ is negative. In the case $\kappa L\in(-1,0)$ this solution is real whereas the case $\kappa L<-1$ yields a purely imaginary wave number which corresponds to a negative energy, referred to as a bound state in the following. The two types of states are continuously connected by the zero energy solution
		\begin{equation}
			\psi_0^{(\mathrm{s})}(x)=\sqrt{\frac{3}{2L^3}}(|x|-L)
		\end{equation}
	valid for $\kappa L=-1$. In the limit $L\to\infty$ the state $\psi_0^{(\mathrm{s})}$ will always be bound irrespective of the value of $\kappa$.
		
	The exact propagator for the confined system can now be written as
		\begin{equation}
			\begin{split}
				K_{\lfloor\delta\rfloor}(x',x,t)
				=&
				\sum_{n=1}^\infty \eh{-\frac{\i\hbar t}{2m}(k_n^{(\mathrm{a})})^2}\psi_n^{(\mathrm{a})}(x')\psi_n^{(\mathrm{a})}(x)
				\\&+
				\sum_{n=0}^\infty \eh{-\frac{\i\hbar t}{2m}(k_n^{(\mathrm{s})})^2}\psi_n^{(\mathrm{s})}(x')\psi_n^{(\mathrm{s})}(x).
			\end{split}
			\label{eqn:finite_propagator}
		\end{equation}
	In the case $\kappa\geq0$ both summations start with $n=1$. In the limit $L\to\infty$ this yields (see Appendix \ref{app:continuum_limit})
		\begin{equation}
			\begin{split}
				K_\delta(x',x,t)
				=&
				\frac{1}{2\pi}\int_{-\infty}^{\infty}\diff{k}\eh{-\frac{\i\hbar t}{2m}k^2}\cos(k(x'-x))
				\\&-
				\frac{1}{2\pi}\int_{-\infty}^{\infty}\diff{k}\eh{-\frac{\i\hbar t}{2m}k^2}\Re\frac{\eh{ik(|x'|+|x|)}}{1+i\frac{k}{\kappa}}
				\\&-
				\Theta(-\kappa)\kappa\eh{\frac{\i\hbar t}{2m}\kappa^2}\eh{\kappa(|x'|+|x|)},
			\end{split}
			\label{eqn:infinite_propagator}
		\end{equation}
	where the last term originates from the separate treatment of the (bound) state $\psi_0^{(\mathrm{s})}$ ($\Theta$ is the Heaviside step function). The first integral in Eq.\@ (\ref{eqn:infinite_propagator}) can be evaluated directly by means of a Gaussian (or Fresnel) integral and yields the well-known expression for the free propagator
	\begin{equation}
		K_0(x',x,t)=\freeprop{(x'-x)}.
	\end{equation}
	The second integral in Eq.\@ (\ref{eqn:infinite_propagator}) can be evaluated in the same way after replacing
		\begin{equation}
			\frac{1}{1+\i\frac{k}{\kappa}}=\int_{0}^{\infty}\diff{\epsilon}\eh{-(1+\i\frac{k}{\kappa})\epsilon}.
		\end{equation}
	Finally, the propagator for the Hamiltonian in Eq. (\ref{eqn:schroedinger_eq}) reads
		\begin{equation}
			K_\delta(x',x,t)=K_0(x',x,t)+K_\kappa(x',x,t),
			\label{delta.propagator}
		\end{equation}
	with the deviation from free propagation given by
		\begin{equation}
			\begin{split}
				K_\kappa&(x',x,t)
				=
					\frac{1\mp 1}{2}\kappa\eh{\frac{\i\hbar t}{2m}\kappa^2}\eh{-\kappa(|x'|+|x|)}
				\\&-
				\kappa\sqrt{\frac{m}{2\pi\i\hbar t}}\int_{0}^{\infty}\diff{\epsilon}\eh{-\kappa\epsilon}\eh{-\frac{m}{2i\hbar t}(|x'|+|x|\pm\epsilon)^2}.
			\end{split}
			\label{eqn:kappaprop}
		\end{equation}
	Here, we identified $\kappa=|\kappa|$ and the upper (lower) signs stand for a repulsive (attractive) potential. This result generalizes the propagator found in \cite{manoukian1989,gaveau1986}, which is restricted to the repulsive case, and is equivalent to the result from path integral approaches \cite{bauch1985}. Note that the second term in Eq.\@ (\ref{eqn:kappaprop}) can be written as
		\begin{equation}
			-\int_{0}^{\infty}\diff{\epsilon}\kappa\eh{-\kappa\epsilon}K_0(-|x'|,|x|\pm\epsilon,t),
		\end{equation}
	which is closely related to the correction $-K_0(-x',x,t)$ obtained for a Dirichlet boundary condition ($\kappa\to\infty$) at $x=0$ \cite{baltes1976}. It thus can be interpreted as the propagation from $x$ to $x'$ via the $\delta$-potential, taking a detour or shortcut of length $\epsilon$ weighted with the density $\kappa\eh{-\kappa\epsilon}$.
\subsection{\boldmath The DOS for a billiard with a $\delta$-barrier}
	The DOS of a $d$-dimensional system can be written as the inverse Laplace transform of the trace of the propagator \cite{hummel2013}:
		\begin{equation}
			\rho(E)=\invLaplace{\beta}{\int\Diff{d}{x}K(\vect{x},\vect{x},t=-\i\hbar\beta)}{E}.
			\label{eqn:inv_lapl}
		\end{equation}
	Let $\Omega$ be the configuration space of a $d$-dimensional billiard with a classically impenetrable thin barrier inside the volume which can be approximated by a $\delta$-potential along a $(d-1)$-dimensional smooth manifold. By only taking into account short-time propagation inside the billiard we can locally approximate the barrier by $(d-1)$-dimensional planes and thus treat the coordinate perpendicular to the barrier as independent from the remaining $d-1$ tangential coordinates. In this case the local approximation for the propagator can be written as
		\begin{align}
			K^{(d)}(\vect{x'},\vect{x},t)
			=&
			K_0^{(d-1)}(\vect{x}_\parallel',\vect{x}_\parallel,t) K_\delta(x_\perp',x_\perp,t)\nonumber
			\\=&
			K_0^{(d)}(\vect{x}',\vect{x},t)
			\\&+
			K_0^{(d-1)}(\vect{x}_\parallel',\vect{x}_\parallel,t) K_\kappa(x_\perp',x_\perp,t)\nonumber.
		\end{align}
	Here, $K_0^{(d)}(\vect{x'},\vect{x},t)$ is the $d$-dimensional free propagator and $x_\perp$ and $\vect{x}_\parallel$ are the coordinates perpendicular and tangential to the barrier. The trace is now calculated assuming that the integration of the perpendicular direction converges rapidly and is thus independent of the position at the barrier. By introducing the interaction strength in units of energy,
		\begin{equation}
			\mu=\frac{\hbar^2\kappa^2}{2m},
		\end{equation}
	this yields
		\begin{equation}
			\begin{split}
				\int_{\Omega}\Diff{d}{x}& K(\vect{x},\vect{x},t=-\i\hbar\beta)
				=
					V_\Omega\Big(\frac{m}{2\pi\hbar^2\beta}\Big)^{\frac{d}{2}}
				\\&+
					\frac{S_\delta}{2}\Big(\frac{m}{2\pi\hbar^2\beta}\Big)^{\frac{d-1}{2}}
				\\& \quad \times
					\big(-1\pm\eh{\mu\beta}\erfc(\sqrt{\mu\beta})+(1\mp 1)\eh{\mu\beta}\big)
			\end{split}
			\label{eqn:general_trace}
		\end{equation}
	with the $d$-dimensional volume $V_\Omega$ of the configuration space $\Omega$ and the surface $S_\delta$ of the barrier. The DOS is then given by the inverse Laplace transform of Eq.\@ (\ref{eqn:general_trace}). For $d=1$ we have to set $S_\delta=1$ and the DOS is given by
		\begin{equation}\begin{split}
			\rho(E)=& V_\Omega\Big(\frac{m}{2\pi\hbar^2}\Big)^\frac{1}{2}\frac{\Theta(E)}{\sqrt{\pi E}}
				-\frac{1}{2}\delta(E)
			\\&\pm
				\frac{1}{2\pi}\sqrt{\frac{\mu}{E}}\frac{\Theta(E)}{E+\mu}
				+\frac{1\mp 1}{2}\delta(E+\mu).
		\end{split}\end{equation}
	In all other cases the result is
	\begin{widetext}
		\begin{equation}
			\begin{split}
				\rho(E)
				=&
					V_\Omega\left(\frac{m}{2\pi\hbar^2}\right)^\frac{d}{2}\frac{E^\frac{d-2}{2}}{\Gamma(\frac{d}{2})}\Theta(E)
				-
					\frac{S_\delta}{2}\left(\frac{m}{2\pi\hbar^2}\right)^\frac{d-1}{2}\frac{E^\frac{d-3}{2}}{\Gamma(\frac{d-1}{2})}\Theta(E)
				\\&\pm
					\frac{S_\delta}{2}\left(\frac{m}{2\pi\hbar^2}\right)^\frac{d-1}{2}
					\left\{\invLaplace{\beta}{\frac{1}{\beta^\frac{d-1}{2}}\eh{\mu\beta}\erfc(\sqrt{\mu\beta})}{E}
				+
					(1\mp 1) \frac{(E+\mu)^\frac{d-3}{2}}{\Gamma(\frac{d-1}{2})}\Theta(E+\mu)\right\}.
			\end{split}
			\label{eqn:delta-DOS}
		\end{equation}
	\end{widetext}
	A closed formula for the inverse Laplace transform
	\begin{equation}
		\invLaplace{\beta}{\frac{1}{\beta^\frac{d-1}{2}}\eh{\mu\beta}\erfc(\sqrt{\mu\beta})}{E}
	\end{equation}
	for arbitrary dimensions $d>1$ is given in Appendix \ref{app:inverse_laplace}.
	
	For varying strength of the $\delta$-potential along the surface of the barrier, i.e., $\kappa=\kappa(\vect{x})$, the surface $S_\delta$ has to be replaced by the integral operator $\int_{S_\delta}\Diff{d-1}{x}$. Note that the boundary conditions at the boundary $\partial\Omega$ of the billiard are not yet included and the approximation of a flat barrier may fail near the boundary. Furthermore, the above approximation does not include curvature corrections and contains only information on the smooth part of the DOS. The result could, in principle, be improved by using periodic orbit theory following \cite{sieber1995}, but this would be at the expense of generality.
	
	Now consider a different setup without a $\delta$-barrier inside the billiard but with Robin- (or mixed) boundary conditions
		\begin{equation}
			\frac{\partial}{\partial x_\perp}\psi(\vect{x})\bigg\vert_{\vect{x}_s}=\kappa\psi(\vect{x}_s),\quad \vect{x}_s\in\partial\Omega.
		\end{equation}
	In 1D this is equivalent to a $\delta$-potential $(\hbar^2\kappa/m)\,\delta(x_s)$ at the surface (i.e., the end points of the line segment) while only allowing for solutions symmetric to the endpoints in a coordinate space extended beyond the latter. This means that in the approximation of a locally flat surface of the boundary we only have to replace the propagator $K_\delta$ in the above derivation by its symmetry-projected equivalent
		\begin{equation}
			\begin{split}
				K_\delta^+(x',x,t)
				=&
					K_\delta(x',x,t)+K_\delta(-x',x,t)
				\\=&
					K_0(x',x,t)+K_0(-x',x,t)
				\\&+
					2K_\kappa(x',x,t)
			\end{split}
			\label{eqn:sym_prop}
		\end{equation}
	while taking the trace perpendicular to the boundary to one side only. One can easily see that this only adds an additional surface term
		\begin{equation}
			\frac{S_\delta}{4}\left(\frac{m}{2\pi\hbar^2}\right)^\frac{d-1}{2}\frac{E^\frac{d-3}{2}}{\Gamma(\frac{d-1}{2})}\Theta(E)
			\label{eqn:neumann_addition}
		\end{equation}
	to Eq.\@ (\ref{eqn:delta-DOS}), which corresponds to a Neumann boundary condition, while the other terms remain unchanged. This is due to the fact that the Robin boundary condition is equivalent to a $\delta$-potential on the surface combined with a Neumann boundary condition in the sense of reflection symmetry in the extended space. The above result reproduces the first term in the expansion derived in \cite{balian1971}. Note that our approach also comprises the two-dimensional (2D) case, which was treated separately in \cite{balian1971}. This will be essential in the next section, where we will use the results to derive the Weyl expansion for an interacting system.
	
\section{Two particles on a line segment}
	\subsection{Configuration space}
\begin{figure}
		\includegraphics{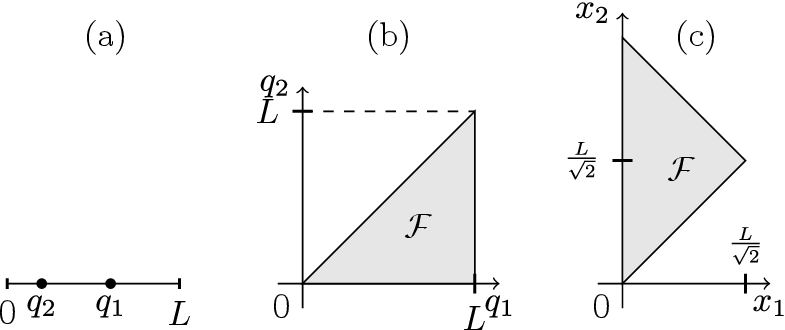}
		\caption{Equivalence of a (a) 1D two-particle system to a (b) 2D one-particle system with fundamental domain $\mathcal{F}$; (c): fundamental domain after a rotation of the coordinates.}
		\label{fig:fundamental_domain}
	\end{figure}

	In this section we consider two identical particles on a line with Dirichlet boundary conditions at $q=0$ and $q=L$ [two particles in a box, see Fig.\@ \ref{fig:fundamental_domain}(a)] as an idealized model for confined particles. The particles shall interact only when they are at the same point, which is realized by a $\delta$-potential. Furthermore we restrict ourselves to either bosons with zero spin or fermions with spin $1/2$. In order to shorten notation, we use imaginary time and choose scaled units, i.e.,
	\begin{equation}
	\beta=\frac{it}{\hbar}\qquad\text{and}\qquad\frac{\hbar^2}{2m}=1.
	\label{eqn:units}
	\end{equation}
	Inside the configuration space $\Omega=\{\vect{q}\in\R^2\mid 0<q_i<L\}$ the Hamiltonian is given by
		\begin{equation}
			\hamilton(q_1,q_2)=-\frac{\partial^2}{\partial q_1^2}-\frac{\partial^2}{\partial q_2^2}+\sqrt{8}\ktilde\,\delta(q_1-q_2).
		\end{equation}
	After introducing relative and center of mass coordinates
		\begin{equation}
			x_1=\frac{1}{\sqrt{2}}(q_1-q_2),\quad x_2=\frac{1}{\sqrt{2}}(q_1+q_2)
		\end{equation}
	the Hamiltonian reads
		\begin{equation}
			\hamilton(x_1,x_2)=-\frac{\partial^2}{\partial x_1^2}-\frac{\partial^2}{\partial x_2^2}+2\ktilde\,\delta(x_1).
			\label{eqn:Hamiltonian}
		\end{equation}
	For bosons, the wave functions must be symmetric with respect to particle exchange and, in the case of fermions, an interaction can only occur if the particles have different spin, i.e., the wave functions must be symmetric, too. We can therefore restrict ourselves to bosons. For $\ktilde=0$ the system of two indistinguishable particles is equivalent to a system of one quasi-particle of the same mass in the fundamental domain $\mathcal{F}=\{\vect{q}\in\Omega\mid q_1\geq q_2\}$ while requiring a Neumann boundary condition on the line $q_1=q_2$ \cite{hummel2013}. In the interacting case the same arguments yield the very same equivalence but with a Robin boundary condition
		\begin{equation}
			\frac{\partial}{\partial x_1}\psi(x_1,x_2)\bigg\vert_{x_1=0}=\ktilde\psi(0,x_2)
		\end{equation}
	on the symmetry line instead, as illustrated in Fig.\@ \ref{fig:fundamental_domain}(b),(c).
\subsection{Smooth part of the DOS}
\begin{figure}
		\centering
		\includegraphics{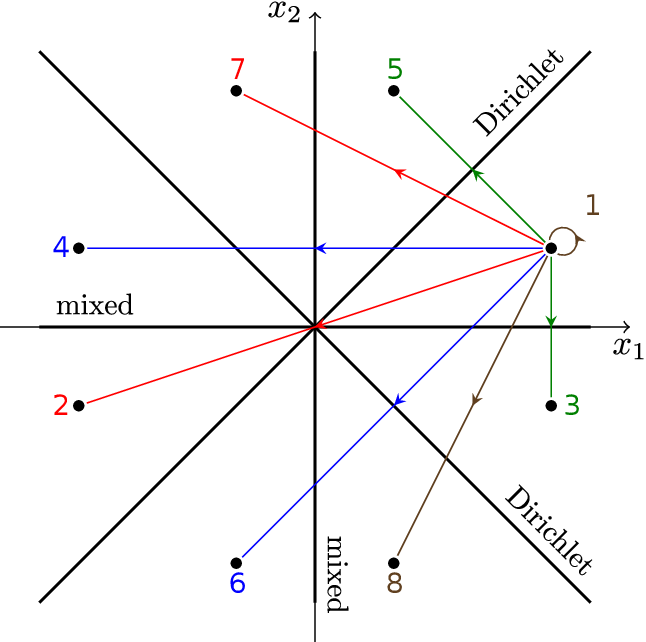}
		\caption{Interpretation of the different terms in the propagator (\ref{eqn:prop_corner}) for $\vect{x}'=\vect{x}$ numbered in order of appearance.}
		\label{fig:prop_corner_1}
	\end{figure}

\begin{figure}
		\centering
		\includegraphics{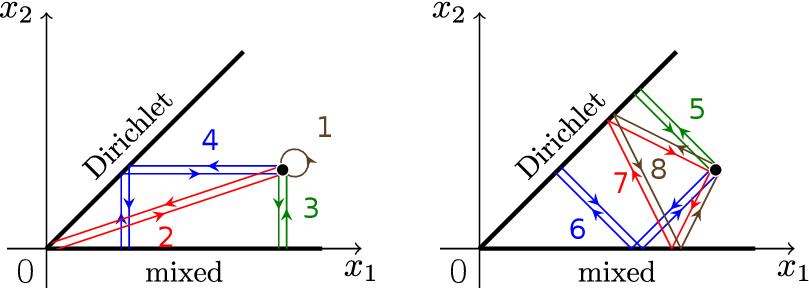}
		\caption{Interpretation of the terms 1--8 in the propagator (\ref{eqn:prop_corner}) associated with the corner as multiple reflections. The trajectories in the right figure include a single reflection at the ray with Dirichlet boundary condition, which results in a minus sign in Eq.\@ (\ref{eqn:prop_corner}).}
		\label{fig:prop_corner_2}
	\end{figure}

	Using the above equivalence of two particles on a line to one quasi-particle in the fundamental domain we can calculate the DOS directly from Eqs.\@ (\ref{eqn:delta-DOS}) and (\ref{eqn:neumann_addition}):
		\begin{equation}
			\begin{split}
				\rho(E)
				=&
				\frac{L^2}{8\pi}\Theta(E)
				-
				\frac{(2+\sqrt{2})L}{8\pi}\frac{\Theta(E)}{\sqrt{E}}
				\\&\pm
				\frac{\sqrt{2}L}{4\pi}\frac{\Theta(E)}{\sqrt{E+\mutilde}}
				+
				\tkappa\frac{\sqrt{2}L}{2\pi}\frac{\Theta(E+\mutilde)}{\sqrt{E+\mutilde}}
				\\&+
				\rho_c(E),
				\label{eqn:DOS_no_corners}
			\end{split}
		\end{equation}
	with $\mutilde=\ktilde^2$ in scaled units (\ref{eqn:units}). We introduced the abbreviation $\tkappa=(1\mp 1)/2$ here in order to shorten notation. The last term $\rho_c(E)$ represents the contributions coming from the corners in the fundamental domain. Here we need the contribution from a $\pi/4$ corner with Dirichlet and Robin boundary conditions along the rays. The exact propagator for such a corner can be derived from the propagator for a $\pi/2$ corner with Robin boundary conditions on the axes, 
		\begin{equation}
			K_{\frac{\pi}{2}}(\vect{x}',\vect{x},t)=K_\delta^+(x_1',x_1,t)K_\delta^+(x_2',x_2,t).
		\end{equation}
	The Dirichlet boundary condition at $x_1=x_2$ can be satisfied by antisymmetrizing $K_{\frac{\pi}{2}}$ with respect to this line. This yields the expression
		\begin{equation}
			\begin{split}
				K_{\frac{\pi}{4}}(\vect{x'},\vect{x},t)
				=&
				\frac{1}{2}\Big[K_{\frac{\pi}{2}}\bm{(}(x_1',x_2'),(x_1,x_2),t\bm{)}
				\\&-K_{\frac{\pi}{2}}\bm{(}(x_2',x_1'),(x_1,x_2),t\bm{)}\Big].
				\label{eqn:red_prop_corners_short}
			\end{split}
		\end{equation}
	The factor $1/2$ was chosen such that the trace can be taken in the first quadrant instead of only integrating the inside of the corner. This is possible due to the symmetry of $K_{\frac{\pi}{2}}$. It is now convenient to write the symmetric propagator (\ref{eqn:sym_prop}) as
		\begin{equation}
			K_\delta^+(x',x,t)=K_0(x',x,t)+K_R(x',x,t)
		\end{equation}
		with
		\begin{equation}
			K_R(x',x,t)=K_0(-x',x,t)+2K_\kappa(x',x,t).
		\end{equation}
	Here, $K_R(x',x,t)$ can be interpreted as the correction to the free propagator representing the propagation from $x$ to the point $x'$ via the boundary with mixed boundary condition. The propagator $K_{\frac{\pi}{4}}$ now takes the form
		\begin{equation}
			\begin{split}
				K_{\frac{\pi}{4}}(\vect{x'},\vect{x},t)
				=&
				\frac{1}{2}\big[K_0(x_1',x_1,t)K_0(x_2',x_2,t)
				\\&+
				K_R(x_1',x_1,t)K_R(x_2',x_2,t)
				\\&+
				K_0(x_1',x_1,t)K_R(x_2',x_2,t)
				\\&+
				K_R(x_1',x_1,t)K_0(x_2',x_2,t)
				\\&
				-K_0(x_2',x_1,t)K_0(x_1',x_2,t)
				\\&-
				K_R(x_2',x_1,t)K_R(x_1',x_2,t)
				\\&-
				K_0(x_2',x_1,t)K_R(x_1',x_2,t)
				\\&-
				K_R(x_2',x_1,t)K_0(x_1',x_2,t)\big].
				\label{eqn:prop_corner}
			\end{split}
		\end{equation}
	The interpretation of the different terms for $\vect{x}'=\vect{x}$ is shown in Figs.\@ \ref{fig:prop_corner_1} and \ref{fig:prop_corner_2}.
	Tracing the above propagator will lead to different contributions to the DOS. The first term can be identified as a volume term, and the third and fourth can be combined to a surface term corresponding to mixed boundary conditions. The surface terms for the Dirichlet boundary are given by the fifth and part of the sixth term. All these contributions have to be dropped to get the parts that arise only from the corner:
		\begin{equation}
			\begin{split}
				\overline{K}_{\frac{\pi}{4}}(\vect{x}',\vect{x},t)
				=&
				\frac{1}{2}\big[K_R(x_1',x_1,t)K_R(x_2',x_2,t)
				\\&-
				K_R(x_2',x_1,t)K_R(x_1',x_2,t)
				\\&+
				K_0(-x_2',x_1,t)K_0(-x_1',x_2,t)
				\\&-
				K_0(x_2',x_1,t)K_R(x_1',x_2,t)
				\\&-
				K_R(x_2',x_1,t)K_0(x_1',x_2,t)\big].
			\end{split}
			\label{eqn:red_prop_corner}
		\end{equation}
	The inverse Laplace transform (\ref{eqn:inv_lapl}) of the trace of this propagator yields the corner contribution. For the first term in Eq.\@ (\ref{eqn:red_prop_corner}) the trace can be calculated separately for each coordinate and the inverse Laplace transform can be calculated as a convolution of the density
		\begin{equation}
			\begin{split}
				&\invLaplace{\beta}{\int_{0}^{\infty}\diff{x} K_R(x,x,t)}{E}
				\\&=
				\invLaplace{\beta}{\left(-\frac{1}{4}\pm\frac{1}{2}\eh{\mutilde\beta}\erfc(\sqrt{\mutilde\beta})+\tkappa\eh{\mutilde\beta}\right)}{E}
				\\&=
				-\frac{1}{4}\delta(E)
				\pm\frac{1}{2\pi}\sqrt{\frac{\mutilde}{E}}\frac{\Theta(E)}{E+\mutilde}
				+\tkappa\delta(E+\mutilde)
			\end{split}
		\end{equation}
	with itself. The remaining four terms of the propagator (\ref{eqn:red_prop_corner}) are traced as a whole because most of the integrals do not have to be evaluated as they cancel mutually after some elementary manipulations. Altogether, the $\pi/4$ corner contribution is
		\begin{align}
			\rho&_{\frac{\pi}{4}}(E)
			=
			\frac{5}{32}\delta(E)
			+
			\frac{1}{4\pi}\sqrt{\frac{\mutilde}{E+\mutilde}}\frac{\Theta(E)}{E+2\mutilde}\nonumber
			\\&\mp
			\frac{1}{8\pi}\sqrt{\frac{\mutilde}{E}}\frac{\Theta(E)}{E+\mutilde}
			\mp
			\frac{1}{4\pi}\sqrt{\frac{2\mutilde}{E}}\frac{\Theta(E)}{E+2\mutilde}
			\\&-
			\tkappa
			\left[
				\frac{1}{4}\delta(E+\mutilde)
				+
				\frac{1}{2\pi}\sqrt{\frac{\mutilde}{E+\mutilde}}\frac{\Theta(E+\mutilde)}{E+2\mutilde}
			\right].\nonumber
			\label{eqn:DMEcke}
		\end{align}
	Note that all the $\mutilde$-dependent expressions give multiples of $\delta(E)$ in the limit $\mutilde\to 0$ (Neumann case), as expected. Combining this result with the contribution from a $\pi/2$ Dirichlet-Dirichlet corner (see, e.g.\@, \cite{eckhardt1988}) we finally obtain the entire DOS for the system
	\begin{widetext}
		\begin{equation}
			\begin{split}
				\rho(E)
				=&
				\frac{L^2}{8\pi}\Theta(E)
				-
				\frac{(2+\sqrt{2})L}{8\pi}\frac{\Theta(E)}{\sqrt{E}}
				\pm
				\frac{\sqrt{2}L}{4\pi}\frac{\Theta(E)}{\sqrt{E+\mutilde}}
				+
				\tkappa\frac{\sqrt{2}L}{2\pi}\frac{\Theta(E+\mutilde)}{\sqrt{E+\mutilde}}
				\\&+
				\frac{3}{8}\delta(E)
				\mp
				\frac{1}{4\pi}\sqrt{\frac{\mutilde}{E}}\frac{\Theta(E)}{E+\mutilde}
				\mp
				\frac{1}{2\pi}\sqrt{\frac{2\mutilde}{E}}\frac{\Theta(E)}{E+2\mutilde}
				+
				\frac{1}{2\pi}\sqrt{\frac{\mutilde}{E+\mutilde}}\frac{\Theta(E)}{E+2\mutilde}
				\\&-
				\tkappa
				\left[
					\frac{1}{2}\delta(E+\mutilde)
					+
					\frac{1}{\pi}\sqrt{\frac{\mutilde}{E+\mutilde}}\frac{\Theta(E+\mutilde)}{E+2\mutilde}
				\right].	
			\end{split}
				\label{eqn:DOS_with_corners}
		\end{equation}
	\end{widetext}
	The above result can be represented in a shorter form if we consider positive energies only. Then the cases of an attractive and a repulsive interaction only differ in the overall sign of the $\tilde{\mu}$-dependent corner corrections:
		\begin{equation}
			\begin{split}
				\rho_+(E)
				=&
				\frac{L^2}{8\pi}
				-
				\frac{(2+\sqrt{2})L}{8\pi}\frac{1}{\sqrt{E}}
				+
				\frac{\sqrt{2}L}{4\pi}\frac{1}{\sqrt{E+\mutilde}}
				\\&\mp
				\left[
					\frac{1}{4\pi}\sqrt{\frac{\mutilde}{E}}\frac{1}{E+\mutilde}
					+
					\frac{1}{2\pi}\sqrt{\frac{2\mutilde}{E}}\frac{1}{E+2\mutilde}
				\right.
				\\&\qquad-\left.
					\frac{1}{2\pi}\sqrt{\frac{\mutilde}{E+\mutilde}}\frac{1}{E+2\mutilde}
				\right] +\frac{3}{8}\delta(E).
			\end{split}
			\label{eqn:DOS_positive_energy}
		\end{equation}
	Note that the corner corrections are monotonous and nonzero for $E>0$ but vanish for $E\to\infty$. This means that the corner corrections are most important for energies near the ground state or, in the case of an attractive interaction, close to $E=0$.
	
	In the regime $-\mutilde<E<0$ the DOS is conveniently expressed in terms of the excitation energy $E^*=E+\mutilde$:
		\begin{equation}
			\rho_{-}(E^*)=\frac{\sqrt{2}L}{2\pi}\frac{\Theta(E^*)}{\sqrt{E^*}}-\frac{1}{2}\delta(E^*)-\frac{1}{\pi}\sqrt{\frac{\mutilde}{E^*}}\frac{\Theta(E^*)}{E^*+\mutilde}.
			\label{eqn:DOS_negative_energy}
		\end{equation}
	The first two terms correspond exactly to the DOS for one particle of mass $2m$ on a line with Dirichlet boundary conditions at the end points: for very strong interaction the two particles behave as one single particle. The third term	represents the interplay of boundary reflections and the interaction. If integrated from $-\mu$ to $0$ it reduces the level counting function by exactly $1/2$ bound state.
\subsection{Comparison with numerical calculations}	
	The comparison with the exact quantum mechanical DOS will be done using the level counting function
		\begin{equation}
		\mathcal{N}(E)=\int_{-\infty}^{E}\diff{E'}\rho(E').
		\end{equation}
	In all plots $E$ and $\tilde{\mu}$ will be given in units of $1/L^2$ ($\tilde{\kappa}$ in units of $1/L$) which is equivalent to setting $L=1$ in $\mathcal{N}$.
		\begin{figure}
			\includegraphics[width=\linewidth]{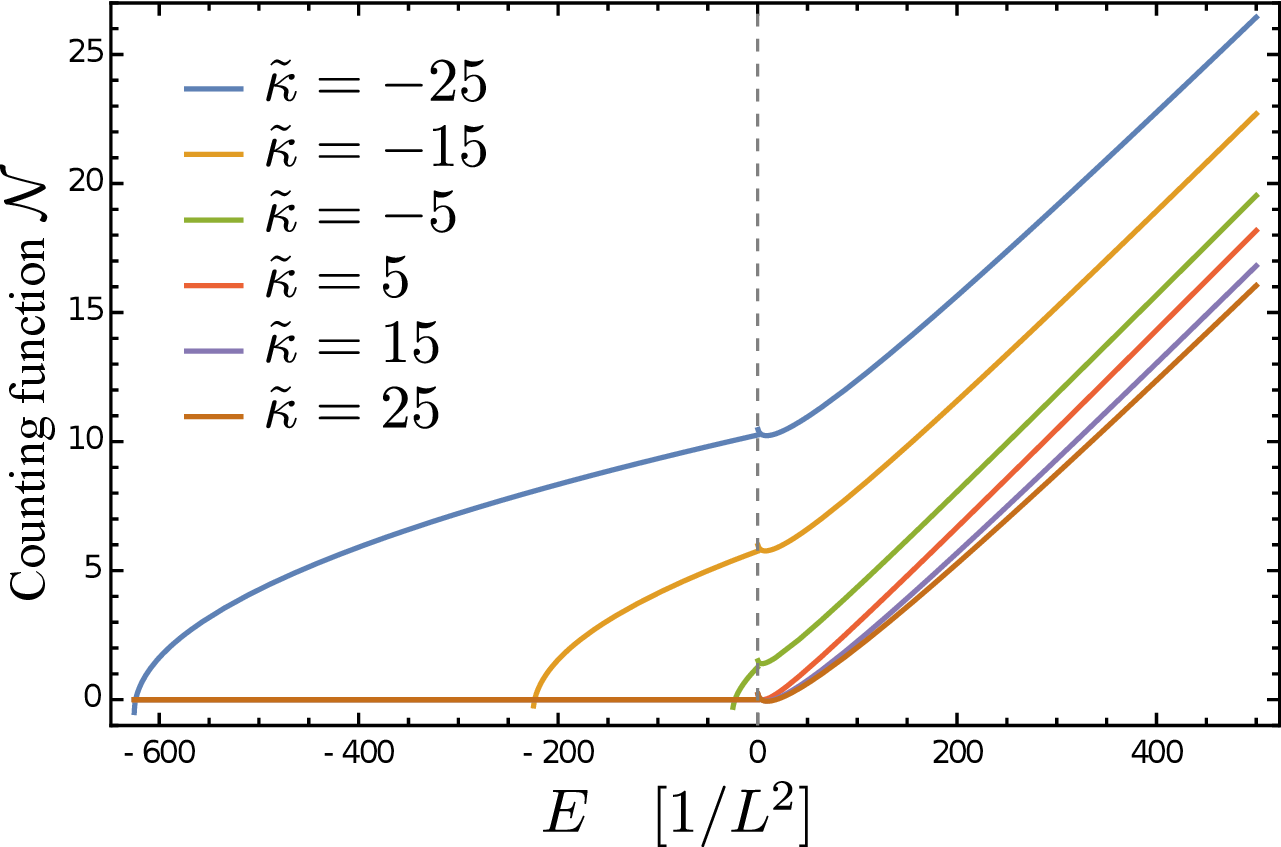}
			\caption{Semiclassical level counting function for $L=1$ and values of $\ktilde$ from $-25$ (left) to $25$ (right) in steps of $\Delta \tilde{\kappa}=10$.}
			\label{fig:smooth_DOS}
		\end{figure}
	In Fig.\@ \ref{fig:smooth_DOS} the semiclassical level counting function is plotted for $-25\leq \ktilde\leq 25$ in steps of $\Delta\ktilde=10$. For $E<0$ the three curves representing the attractive cases $\kappa=-25,-15,-5$ resemble 1D single-particle counting functions.
	
\begin{figure}
	\centering
		\includegraphics{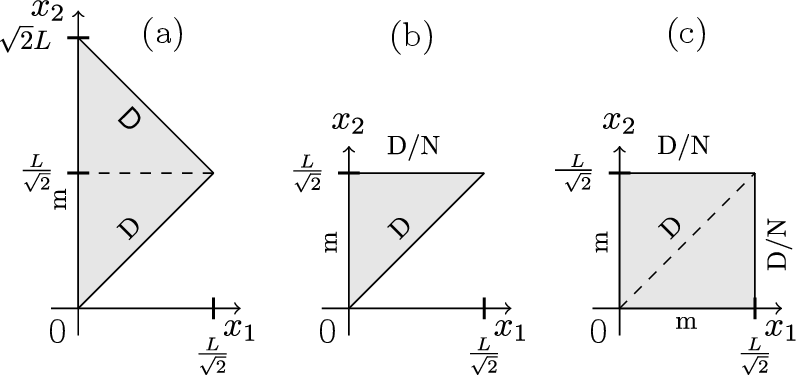}
		\caption{Fundamental domain and equivalent systems. The letters stand for mixed (m), Neumann (N) and Dirichlet (D) boundary conditions.}
		\label{fig:exact_sols}
	\end{figure}

	The exact solutions of the Schr\"{o}dinger equation for the Hamiltonian (\ref{eqn:Hamiltonian}) can be found by using that the system is symmetric with respect to the line $x_2=L/\sqrt{2}$. This means that we can determine a full set of energy eigenstates that are either symmetric or antisymmetric to that line [see Fig.\@ \ref{fig:exact_sols}(a)]. Except for normalization, this is equivalent to considering only the lower part of the fundamental domain while requiring either Neumann or Dirichlet boundary conditions at the symmetry line [see Fig.\@ \ref{fig:exact_sols}(b)]. The Dirichlet boundary condition at the line $x_1=x_2$ corresponds to the antisymmetric solutions of the extended system shown in Fig.\@ \ref{fig:exact_sols}(c).
	Moreover, the normalization constants for the original and the extended system are the same. The solutions can be written straightforwardly as antisymmetrized products of 1D wave functions.
		\begin{equation}
			\begin{split}
				\psi_{mn}^\text{D}
				=&
				\frac{1}{2}\left[\psi_m^\text{D}(x_1)\psi_n^\text{D}(x_2)-\psi_m^\text{D}(x_2)\psi_n^\text{D}(x_1)\right],
				\\&\quad k_m,k_n\in \text{Spec}^\text{D},
				\quad 0\leq m<n,
				\\
				\psi_{mn}^\text{N}
				=&
				\frac{1}{2}\left[\psi_m^\text{N}(x_1)\psi_n^\text{N}(x_2)-\psi_m^\text{N}(x_2)\psi_n^\text{N}(x_1)\right],
				\\&\quad k_m,k_n\in \text{Spec}^\text{N},
				\quad 0\leq m<n.
			\end{split}
			\label{eqn:exact_sols}
		\end{equation}
	Here, the 1D wave functions and the sets $\text{Spec}^{\mathrm{N/D}}$ are defined by
		\begin{align*}
			\begin{split}
				\psi_n^\text{D}(x)
				&=
				A_n^\text{D}
				\sin\left(k_n(x-d)\right),
				\\
				\text{Spec}^\text{D}
				&=
				\left\{k_n\mid k_n d=n\pi-\arctan\left(\frac{k_n}{\ktilde}\right),n\in\N_0\right\}
				\\&\qquad\cup
				\left\{k_0=\i \tilde{k}_0 \mid \tilde{k}_0=-\ktilde\tanh(\tilde{k}_0d),\tilde{k}_0>0\right\},
			\end{split}
			\\
			\begin{split}
				\psi_n^\text{N}(x)
				=&
				A_n^\text{N}\cos\left(k_n(x-d)\right),
				\\
				\text{Spec}^\text{N}
				=&
				\left\{k_n\mid k_n d=n\pi-\frac{\pi}{2}-\arctan\left(\frac{k_n}{\ktilde}\right),n\in\N \right\}
				\\&\qquad\cup
				\left\{k_0=\i \tilde{k}_0\mid \tilde{k}_0=-\ktilde\coth(\tilde{k}_0 d),\tilde{k}_0>0\right\},
			\end{split}
		\end{align*}
	where $k_0\in \text{Spec}^\text{D}$ is chosen either positive or purely imaginary with positive imaginary part depending on the value of $\kappa d$ with $d=L/\sqrt{2}$.
	\begin{figure}
		\includegraphics[width=\linewidth]{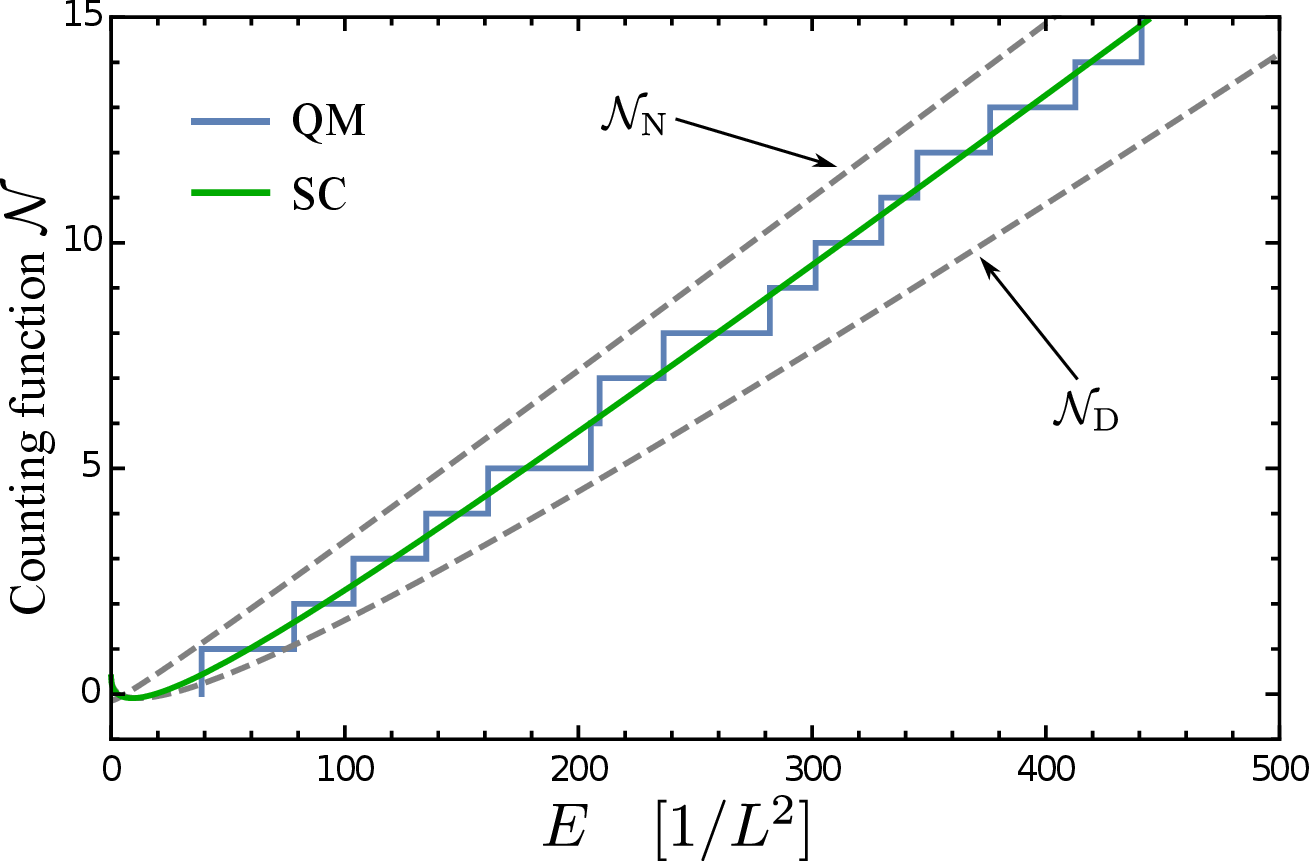}
		\caption{The quantum mechanical (staircase) and semiclassical level counting functions for $\ktilde=10$ (repulsive case) and $L=1$. The functions $\mathcal{N}_\text{N}$ and $\mathcal{N}_\text{D}$ represent the limits $\ktilde\to 0$ and $\ktilde\to\infty$, respectively.}
		\label{fig:DOS_comparison_repulsive}
	\end{figure}
	\begin{figure}
		\includegraphics[width=\linewidth]{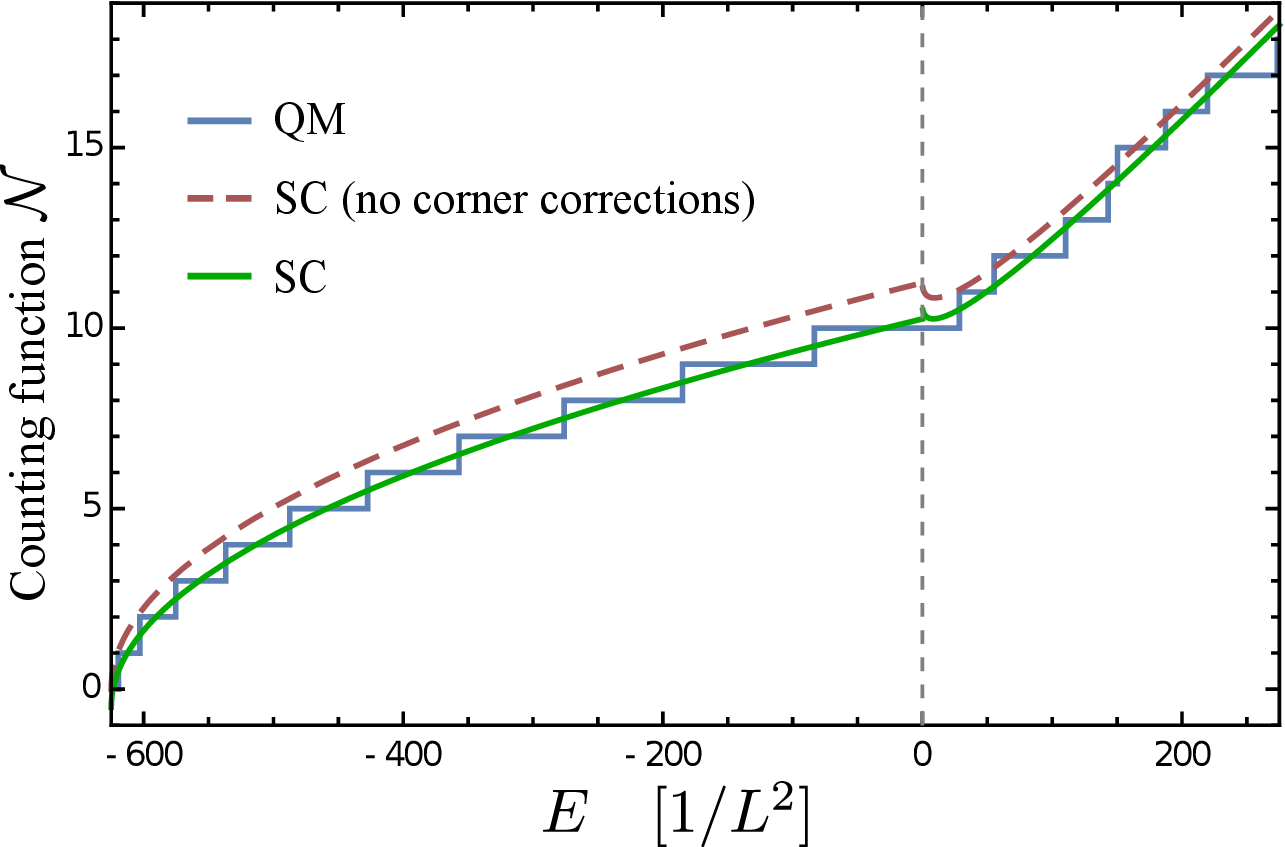}
		\caption{The quantum mechanical (staircase) and semiclassical level counting functions $\mathcal{N}(E)$ for $\ktilde=-25$ (attractive case).}
		\label{fig:DOS_comparison_attractive}
	\end{figure}
	The energy eigenvalues of the system are now given as $E_{mn}=k_m^2+k_n^2$ with $k_m,k_n$ either both in $\text{Spec}^\text{D}$ or both in $\text{Spec}^\text{N}$. The energies have been calculated numerically and the comparison to the smooth level counting function is shown in Fig.\@ \ref{fig:DOS_comparison_repulsive} for the repulsive case ($\ktilde=10$) and in Fig.\@ \ref{fig:DOS_comparison_attractive} for the attractive case ($\ktilde=-25$).
	
	The value $\ktilde=10$ for the repulsive case has been chosen such that the resulting level counting function lies well in between the two limits $\ktilde=0$ and $\ktilde\to\infty$ corresponding to non-interacting bosons and fermions (denoted as $\mathcal{N}_\text{N}$ and $\mathcal{N}_\text{D}$  in Fig.\@ \ref{fig:DOS_comparison_repulsive}). For $\ktilde\geq 0$ numerical calculations show that the ratio of corner corrections in the DOS to the full semiclassical DOS at the ground state has a maximum of about $8\%$ for $\ktilde\approx 5.814$ which is very close to the value at which the equation $E_0=\ktilde^2 $ for the ground state energy holds ($E_0$ varies smoothly from $E_0=2\pi^2$ for $\ktilde=0$ to $E_0=5\pi^2$ for $\ktilde=\infty$). As this ratio decreases with the energy the corner corrections can be neglected in the DOS for high energies. This holds also true for the level counting function, where they decrease rapidly from $3/8$ to $-1/8$ as the energy increases.
	
	Figure 7 (attractive case) shows that the corner corrections are very important for negative energies and result in a shift by one level in the level counting function at $E=0$. In fact the corner corrections give a contribution that nearly exactly reproduces the quantum mechanical energy eigenstates. By integrating Eq.\@  (\ref{eqn:DOS_negative_energy}) to get $\mathcal{N}_{-}(E^*)$ and substituting $k^*=\sqrt{E^*}$ the equation $\mathcal{N}_{-}(k_n^*)+1/2=n$ for $n\in\N$ yields
	\begin{equation}
		k_n^*d=\frac{n\pi}{2}-\arctan\left(\frac{k_n^*}{\ktilde}\right)
		\label{eqn:requatnization}
	\end{equation}
	which gives exactly the allowed real wave numbers in $\text{Spec}^\text{D}$ and $\text{Spec}^\text{N}$. Since the negative quantum-mechanical energy levels are always given as $E=k_0^2+k_n^2$ with $k_0$ purely imaginary and $k_n$ real we can see that the error in the approximation (\ref{eqn:requatnization}) of the eigenenergies lies only in the assumption that the imaginary wave numbers $k_0$ in both $\text{Spec}^\text{D}$ and $\text{Spec}^\text{N}$ are equal to $-\i\ktilde$. This approximation is very good for large absolute values of $\ktilde$ and is still reasonable as long as the requantization with $\rho_{-}$ makes sense, i.e., the ground state energy is negative.
	\begin{figure}
		\includegraphics[width=\linewidth]{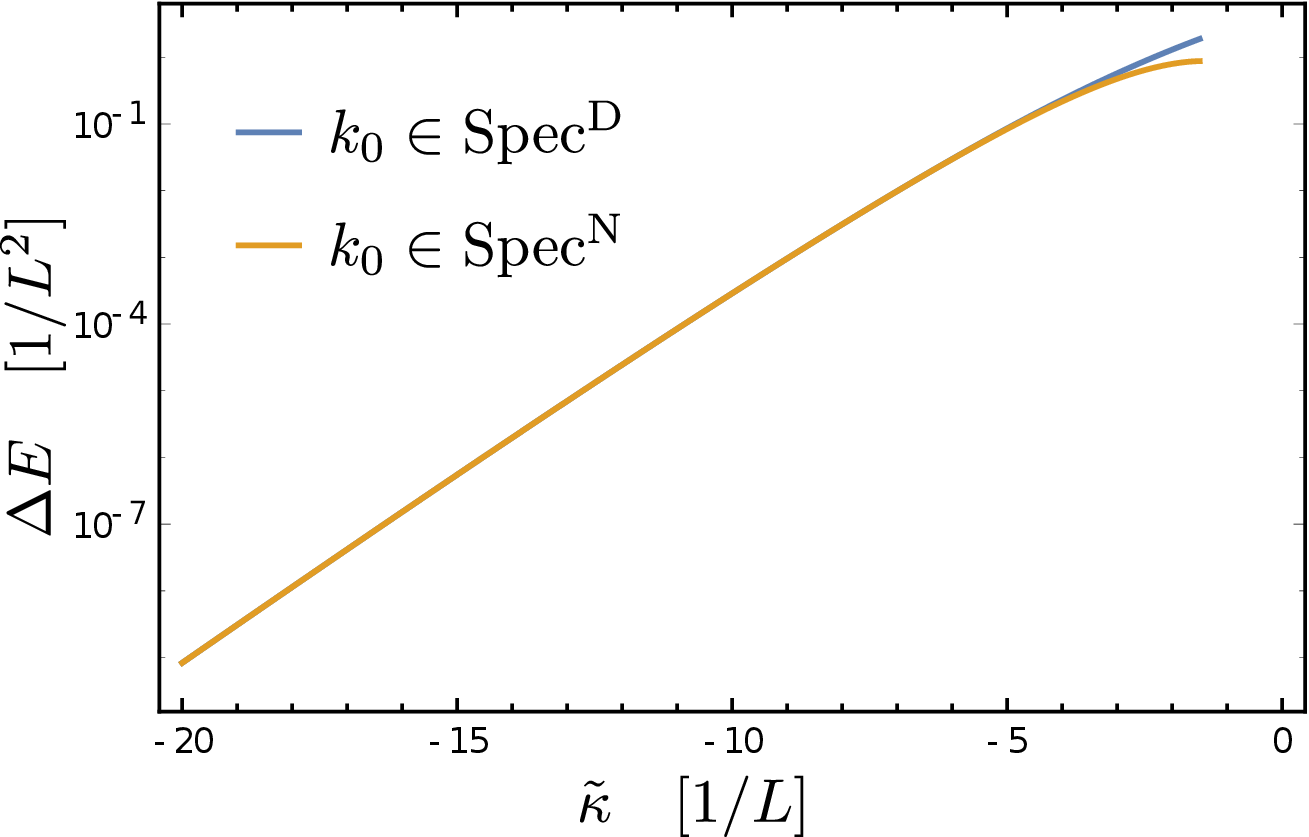}
		\caption{The absolute error $\Delta E_n=|k_0^2-\ktilde^2|$ in the semiclassical approximations for $k_0$ in each set $\text{Spec}^\text{D/N}$.}
		\label{fig:requantization_error}
	\end{figure}
	The error $\Delta E=|k_0^2-\ktilde^2|$ of the semiclassically requantized energies is plotted in Fig.\@ \ref{fig:requantization_error} for $-20<\ktilde<-2$. The ground state energy changes sign at $\ktilde=-(3\pi)/(2\sqrt{2})\approx-3.332$ (semiclassical prediction, while the quantum mechanical result yields $\ktilde\approx-3.286$). At this point the error is less than $0.45/L^2$ which is small compared to the mean level spacing $[\rho_{-}(\tilde{\mu})]^{-1}\approx 19/L^2$. This shows that the corner corrections are essential for $E<0$ and can be used to requantize the system in this regime. Moreover one can use the level counting function directly to find the number of bound states in the system.
	
\section{Conclusion}
	We presented an alternative derivation for the propagator for a $\delta$-potential in Sec.\@ \ref{subsec:propagator}. The resulting expression has a natural interpretation by means of free propagation and hard-wall reflection, taking exponentially weighted detours or shortcuts in the cases of repulsive or attractive interaction, respectively. The propagator can be used to calculate the single-particle DOS for arbitrary shaped billiards of any dimension with mixed boundary conditions and/or $\delta$-barriers inside the volume. Although the calculations do not include curvature terms the formalism is straightforward and allowed us to calculate the contributions from a Dirichlet-Robin $(\pi/4)$-corner to the DOS which, to our knowledge, have not been calculated before. The corrections given by these corners are, as expected, of lowest order in the energy but give very accurate corrections at low positive energies, where the repulsive and the attractive case differ only in an overall sign. In the attractive case the corner corrections allow for the requantization of the system for negative energies, which shows the accuracy of the approximations used in the formalism.
	
	By using the equivalence of the 1D two-particle problem with $\delta$-interactions to the 2D problem in the fundamental domain with mixed boundary conditions we presented a powerful tool for the treatment of few-body systems. Its natural application lies in the approximation of the DOS for an arbitrary number of $\delta$-interacting particles. It is important to note that the scattering and bound states of such a system with infinite volume are known exactly \cite{mcguire1964}. This means that the short-time propagators required for the treatment of higher particle numbers can, in principle, be calculated directly from them. Our approach presented here can be extended to include smooth external potentials, which allowed us to examine the thermodynamics of a system of $\delta$-interacting bosons in a harmonic confinement \cite{hummel2016}.
\begin{acknowledgements}
	B.G.\@ thanks the \textit{Studienstiftung des deutschen Volkes} for support. We	further	acknowledge	support through	the Deutsche Forschungsgemeinschaft.
\end{acknowledgements}

\appendix
\section{Continuum limit of the propagator}
	\label{app:continuum_limit}
	The antisymmetric solutions of the confined system have equidistant $k$'s, i.e.\@, $\Delta k=k_{n+1}-k_n=\pi/L$ for arbitrary $n$. So the first line of Eq.\@ (\ref{eqn:finite_propagator}) is easily verified to have the continuum limit
	\[
		\frac{1}{2\pi}\int_{-\infty}^{\infty}\diff{k}\eh{-\frac{\i\hbar t}{2m}k^2}\big[\cos\bm{(}k(x'-x)\bm{)}-\cos\bm{(}k(x'+x)\bm{)}\big],
	\]
	whereas the second line needs a more careful treatment. Omitting the indices we can write
	\begin{align*}
		\frac{2L}{A^2}\psi(x')\psi(x)=&\cos\bm{(}k(|x'|-|x|)\bm{)}\\&-\cos\bm{(}k(|x'|+|x|-2L)\bm{)}.
	\end{align*}
	Using trigonometric identities and identifying
	\begin{align*}
		\sin(2kL)&=\frac{2\tan(kL)}{1+\tan^2(kL)}=\frac{-2\frac{k}{\kappa}}{1+\left(\frac{k}{\kappa}\right)^2}
		\\
		\cos(2kL)&=\frac{1-\tan^2(kL)}{1+\tan^2(kL)}=\frac{1-\left(\frac{k}{\kappa}\right)^2}{1+\left(\frac{k}{\kappa}\right)^2}
	\end{align*}
	this yields
	\begin{align*}
		\frac{2L}{A^2}\psi(x')\psi(x)=&\cos\bm{(}k(|x'|+|x|)\bm{)}+\cos\bm{(}k(|x'|-|x|)\bm{)}
		\\&-
		2\Re\frac{\eh{ik(|x'|+|x|)}}{1+i\kk},
	\end{align*}
	where the absolute values in the first two arguments can be dropped due to symmetry.
	Now, observing that $A^2=1+O(L^{-1})$ and $\Delta k_n=k_{n+1}-k_n=\pi/L+O(L^{-2})$  for all $n$ as $L$ goes to infinity, the continuum limit can be performed exactly in the same way as for the antisymmetric solutions. The bound state solution of the confined system has an exact limit for the free space, i.e., an exponentially decaying wave function localized at the potential. Summing up all the contributions we get Eq.\@ (\ref{eqn:infinite_propagator}).
\section{Inverse Laplace transformation}
	\label{app:inverse_laplace}
	In Eq.\@ (\ref{eqn:DOS_with_corners}) we need the inverse Laplace transform $f_n(E)$ of functions of the form
		\begin{equation}
			F_n(\beta)=\beta^{-n}\eh{\mu\beta}\erfc(\sqrt{\mu\beta})
		\end{equation}
	with $2n\in\N_0$. To this end one can use the property of the two-sided Laplace transformation
		\begin{equation}
			\Laplace{E}{\int_{-\infty}^{E}\diff{E'}f(E')}{\beta}=\frac{1}{\beta}\Laplace{E}{f(E)}{\beta}
		\end{equation}
	to get $f_n$ recursively from $f_0$ or $f_\frac{1}{2}$. We will take a different approach that yields the full expression. It is straightforward to show
	\begin{align}
		nF_{n+1}(\beta)=\left(\mu-\derivative{\beta}\right)F_n(\beta)-\sqrt{\frac{\mu}{\pi}}\beta^{-(n+\frac{1}{2})}.
	\end{align}
	This can be used to verify by induction the formula
		\begin{equation}
			\begin{split}
				& F_n(\beta)
				=
				\frac{\Gamma(\gamma)}{\Gamma(n)}\left(\mu-\derivative{\beta}\right)^{n-\gamma}F_\gamma(\beta)
				\\&-
				\sqrt{\frac{\mu}{\pi}}
				\sum\limits_{k=1}^{n-\gamma}\frac{\Gamma(n-k)}{\Gamma(n)}\left(\mu-\derivative{\beta}\right)^{k-1}\beta^{-(n-k+\frac{1}{2})},
			\end{split}
		\end{equation}
	where
		\begin{equation}
			\gamma:=
			\begin{cases}
			1 & n\in\N, \\
			\frac{1}{2} & n+\frac{1}{2}\in\N.
			\end{cases}
		\end{equation} 
	Now one can use the linearity of the Laplace transformation and the identity
		\begin{equation}
			\Laplace{E}{-Ef(E)}{\beta}=\derivative{\beta}{}\Laplace{E}{f(E)}{\beta}
		\end{equation}
	to get
		\begin{equation}
			\begin{split}
				f_{n}&(E)
				=
				\frac{\Gamma(\gamma)}{\Gamma(n)}(E+\mu)^{n-\gamma}f_\gamma(E)-
				\\-&
				\sqrt{\frac{\mu}{\pi}}
				\sum\limits_{k=1}^{n-\gamma}\frac{\Gamma(n-k)}{\Gamma(n)\Gamma(n-k+\frac{1}{2})}(E+\mu)^{k-1}E^{n-k-\frac{1}{2}}.
			\end{split}
		\end{equation}
	The functions $f_\gamma$ are given by \cite{abramowitz1965}
		\begin{equation}
			f_\gamma(E)=
			\begin{cases}
				\frac{2}{\pi}\arctan\left(\sqrt{\frac{E}{\mu}}\right)\Theta(E) & \gamma=1, \\
				\frac{1}{\sqrt{\pi}}\frac{\Theta(E)}{\sqrt{E+\mu}} & \gamma=\frac{1}{2}.
			\end{cases}
		\end{equation}

\bibliography{bibliography}

\end{document}